%% file: postNOCI_11-21-25.tex
\documentclass[floatfix,aip,jcp]{revtex4-1}
\usepackage{multirow}
\usepackage{afterpage}
\usepackage{longtable}
\usepackage{xcolor}
\usepackage{array}
\usepackage{amsmath}
\usepackage{physics}
\usepackage{simpler-wick}
\usepackage{pdfcomment}
\usepackage{tikz}
\usepackage[utf8]{inputenc}
\usepackage[T1]{fontenc}
\usepackage{ifthen}
 
\usepackage[top=2.0cm,left=1.5cm,right=1.0cm,bottom=2.25cm]{geometry}
\input{orbital_tickz_def}

\newcounter{mocounter} 
\providecommand{\MO}{abc}
\renewcommand{\MO}
  {\addtocounter{mocounter}{1}\ifthenelse{\equal{\themocounter}{1}}{molecular orbital (MO)}{MO}}
 \providecommand{\MOs}{abc}
\renewcommand{\MOs}
  {\addtocounter{mocounter}{1}\ifthenelse{\equal{\themocounter}{1}}{molecular orbitals (MOs)}{MOs}}
\newcounter{aocounter} 
\providecommand{\AO}{abc}
\renewcommand{\AO}
  {\addtocounter{aocounter}{1}\ifthenelse{\equal{\theaocounter}{1}}{atomic orbital (AO)}{AO}}
 \providecommand{\AOs}{abc}
\renewcommand{\AOs}
  {\addtocounter{aocounter}{1}\ifthenelse{\equal{\theaocounter}{1}}{atomic orbitals (AOs)}{AOs}}
\newcounter{nocounter} 
\providecommand{\NO}{abc}
\renewcommand{\NO}
  {\addtocounter{nocounter}{1}\ifthenelse{\equal{\thenocounter}{1}}{natural orbital (\textit{no})}{\textit{no}}}
 \providecommand{\NOs}{abc}
\renewcommand{\NOs}
  {\addtocounter{nocounter}{1}\ifthenelse{\equal{\thenocounter}{1}}{natural orbitals (\textit{no}s)}{\textit{no}s}}
\newcounter{dodccounter} 
\providecommand{\DODC}{abc}
\renewcommand{\DODC}
  {\addtocounter{dodccounter}{1}\ifthenelse{\equal{\thedodccounter}{1}}{different-orbitals for different-configurations (DODC)}{DODC}}
\newcounter{vbcounter} 
\providecommand{\VB}{abc}
\renewcommand{\VB}
  {\addtocounter{vbcounter}{1}\ifthenelse{\equal{\thevbcounter}{1}}{valence bond (VB)}{VB}}
\newcounter{nocicounter}
\newcounter{nocisdcounter}
\providecommand{\NOCI}{abc}
\renewcommand{\NOCI}
  {\addtocounter{nocicounter}{1}\ifthenelse{\equal{\thenocicounter}{1}}{nonorthogonal configuration interaction (NOCI)}{NOCI}}
    \providecommand{\NOCISD}{abc}
\renewcommand{\NOCISD}
  {\addtocounter{nocicounter}{1}{\addtocounter{nocisdcounter}{1}\ifthenelse{\equal{\thenocicounter}{1}}{nonorthogonal configuration interaction singles and doubles (NOCISD)}{\ifthenelse{\equal{\thenocisdcounter}{1}}{NOCI singles and doubles (NOCISD)}{NOCISD}}}}
\newcounter{scfcounter} 
\providecommand{\SCF}{abc}
\renewcommand{\SCF}
  {\addtocounter{scfcounter}{1}\ifthenelse{\equal{\thescfcounter}{1}}{self-consistent field (SCF)}{SCF}}
\newcounter{mptwocounter} 
\providecommand{\MPTwo}{abc}
\renewcommand{\MPTwo}
  {\addtocounter{mptwocounter}{1}\ifthenelse{\equal{\themptwocounter}{1}}{M{\o}ller-Plesset second-order perturbation theory (MP2)}{MP2}}
\newcounter{mcscfcounter} 
\newcounter{nomcscfcounter} 
\newcounter{ssnomcscfcounter} 
\newcounter{sanomcscfcounter} 
\providecommand{\MCSCF}{abc}
\renewcommand{\MCSCF}
  {\addtocounter{mcscfcounter}{1}\ifthenelse{\equal{\themcscfcounter}{1}}{multiconfigurational self-consistent field (MCSCF)}{MCSCF}}
\providecommand{\NOMCSCF}{abc}
\renewcommand{\NOMCSCF}
  {\addtocounter{mcscfcounter}{1}\addtocounter{nomcscfcounter}{1}\ifthenelse{\equal{\themcscfcounter}{1}}{nonorthogonal multiconfigurational self-consistent field (NOMCSCF)}{\ifthenelse{\equal{\thenomcscfcounter}{1}}{nonorthogonal MCSCF (NOMCSCF)}{NOMCSCF}}}
    \providecommand{\SSNOMCSCF}{abc}
\renewcommand{\SSNOMCSCF}
  {\addtocounter{mcscfcounter}{1}\addtocounter{nomcscfcounter}{1}\addtocounter{ssnomcscfcounter}{1}\ifthenelse{\equal{\themcscfcounter}{1}}{state-specific nonorthogonal multiconfigurational self-consistent field (SS-NOMCSCF)}{\ifthenelse{\equal{\thenomcscfcounter}{1}}{state-specific nonorthogonal MCSCF (SS-NOMCSCF)}{\ifthenelse{\equal{\thessnomcscfcounter}{1}}{state-specifc NOMCSCF (SS-NOMCSCF)}{SS-NOMCSCF}}}}
    \providecommand{\SANOMCSCF}{abc}
\renewcommand{\SANOMCSCF}
  {\addtocounter{mcscfcounter}{1}\addtocounter{nomcscfcounter}{1}\addtocounter{sanomcscfcounter}{1}\ifthenelse{\equal{\themcscfcounter}{1}}{state-average nonorthogonal multiconfigurational self-consistent field (SA-NOMCSCF)}{\ifthenelse{\equal{\thenomcscfcounter}{1}}{state-average nonorthogonal MCSCF (SA-NOMCSCF)}{\ifthenelse{\equal{\thesanomcscfcounter}{1}}{state-average NOMCSCF (SA-NOMCSCF)}{SA-NOMCSCF}}}}
\newcounter{casmaincounter} 
\newcounter{casscfcounter} 
\newcounter{sscasscfcounter} 
\newcounter{sacasscfcounter} 
\providecommand{\CAS}{abc}
\renewcommand{\CAS}
  {\addtocounter{casmaincounter}{1}\ifthenelse{\equal{\thecasmaincounter}{1}}{complete active space (CAS)}{CAS}}
  \providecommand{\CASSCF}{abc}
\renewcommand{\CASSCF}
  {\addtocounter{casmaincounter}{1}\addtocounter{casscfcounter}{1}\ifthenelse{\equal{\thecasmaincounter}{1}}{complete active space self-consistent field (CASSCF)}{\ifthenelse{\equal{\thecasscfcounter}{1}}{CAS self-consistent field (CASSCF)}{CASSCF}}}
    \providecommand{\SSCASSCF}{abc}
\renewcommand{\SSCASSCF}
  {\addtocounter{casmaincounter}{1}\addtocounter{casscfcounter}{1}\addtocounter{sscasscfcounter}{1}\ifthenelse{\equal{\thecasmaincounter}{1}}{state-specific complete active space self-consistent field (SS-CASSCF)}{\ifthenelse{\equal{\thecasscfcounter}{1}}{state-specific CAS self-consistent field (SS-CASSCF)}
{\ifthenelse{\equal{\thesscasscfcounter}{1}}{state-specific CASSCF (SS-CASSCF)}{SS-CASSCF}}}}
    \providecommand{\SACASSCF}{abc}
\renewcommand{\SACASSCF}
  {\addtocounter{casmaincounter}{1}\addtocounter{casscfcounter}{1}\addtocounter{sacasscfcounter}{1}\ifthenelse{\equal{\thecasmaincounter}{1}}{state-average complete active space self-consistent field (SA-CASSCF)}{\ifthenelse{\equal{\thecasscfcounter}{1}}{state-average CAS self-consistent field (SA-CASSCF)}
{\ifthenelse{\equal{\thesacasscfcounter}{1}}{state-average CASSCF (SA-CASSCF)}{SA-CASSCF}}}}
\newcounter{cicounter} 
\providecommand{\CI}{abc}
\renewcommand{\CI}
  {\addtocounter{cicounter}{1}\ifthenelse{\equal{\thecicounter}{1}}{configuration interaction (CI)}{CI}}
\newcounter{icmrcicounter} 
\providecommand{\icMRCI}{abc}
\renewcommand{\icMRCI}
  {\addtocounter{icmrcicounter}{1}\ifthenelse{\equal{\theicmrcicounter}{1}}{internally contracted multireference configuration interaction (ic-MRCI)}{ic-MRCI}}
\newcounter{onepdmcounter} 
\providecommand{\onePDM}{abc}
\renewcommand{\onePDM}
  {\addtocounter{onepdmcounter}{1}\ifthenelse{\equal{\theonepdmcounter}{1}}{one-particle density matrix (1PDM)}{1PDM}}
\newcounter{rasscfcounter} 
\providecommand{\RASSCF}{abc}
\renewcommand{\RASSCF}
  {\addtocounter{rasscfcounter}{1}\ifthenelse{\equal{\therasscfcounter}{1}}{restricted active space self-consistent field (RASSCF)}{RASSCF}}
\newcounter{oscounter} 
\providecommand{\OS}{abc}
\renewcommand{\OS}
  {\addtocounter{oscounter}{1}\ifthenelse{\equal{\theoscounter}{1}}{occupation specific (OS)}{OS}} 

\begin{document}
\title{Natural Excitation Framework for Defining the External Space: Uncontracted and Internally Contracted Multireference Nonorthogonal Wavefunction Theories}
\author{Matheus M. F. Moraes}
\author{Lee M. Thompson}
\email{lee.thompson.1@louisville.edu}
\affiliation{Department of Chemistry, University of Louisville, 2320 South Brook Street, Louisville, KY 40292, USA}

\begin{abstract}
In this communication, we examine new formalisms for the construction of the external space when correlating reference wavefunctions built from nonorthogonal determinant expansions. 
Defining the external space in nonorthogonal approaches is challenging, as every substitution from the reference wavefunction can potentially mix both internal and external configurations. 
As a result, post-nonorthogonal methods are plagued by internal contamination and linear dependencies in the external space, which may lead to correlation double counting that results from overlap of the external and reference spaces. 
Removal of these internal configurations and orthonormalization of the excited space basis can be computationally expensive. In particular, as the excitation operators cannot be subdivided by their action on orbital subspaces, the external space cannot be partitioned into non-overlapping subsets as is possible in orthogonal methods.
To resolve these issues, we propose both uncontracted and internally-contracted approaches based on a natural excitation framework that allows for reduced scaling, more straightforward separation of excitation types that lead to external and internal spaces, and allows for a facile translation of orthogonal methods to a nonorthogonal framework. Several proofs and a numerical demonstration using vanadium monohydride (VH) are provided to illustrate the viability of the proposed approach, using a method-agnostic presentation to highlight the generality of the approach.
\end{abstract}
\maketitle

The development of nonorthogonal wavefunction methods has gained increasing traction in recent years. While \VB{} approaches have long been popular,\cite{Wu.2011} \MO{} based \DODC{} methods that combine features of \MO{} and \VB{} wavefunctions have been the focus of many recent studies. Building on earlier works by Broer,\cite{BROER.1992,Broer:2003et} Fukutome,\cite{Igawa:1988dg,Fukutome:1988ip,Ikawa.1993} and Schlegel\cite{Ayala.1998} that demonstrated the utility of \NOCI{} approaches for describing multireference states, Burton and Thom developed a general protocol valid across all molecular geometries for constructing \NOCI{} wavefunctions from holomorphic \SCF{} solutions.\cite{Burton.2019} Approaches for optimization of all variational parameters (including orbitals) in a \DODC{} framework have also been developed.\cite{Mahler.2021,Miller.2025} Furthermore, \DODC{} expansions are being used to provide compact effective Hamiltonian models in both time-independent and time-dependent frameworks for a broad range of theory developments and applications.\cite{Kempfer-Robertson.2022olv,Thompson.2023,Dong.2024,Saha.2025,Zhu.2023,Lu.2024,Ren.2016,Hettich.2023,Lu.2022,Lu.2022e4s,Zhao.2021bw,Sánchez-Mansilla.2022,Straatsma.2022,Straatsma:2020iv,Kathir:2020ct} However, a remaining challenge for \DODC{} approaches is how post-\NOCI{} methods can be constructed to recover the remaining correlation energy. Addressing this challenge is the central theme of this communication. 

Recovering the dynamic correlation from a \NOCI{} reference was first discussed by Ayala and Schlegel who coupled nonorthogonal symmetry-broken \MPTwo{} wavefunctions using an averaging procedure to ensure Hermicity.\cite{Ayala.1998} This approach was subsequently generalized by Yost and coworkers to deal with general determinant expansions of \SCF{} solutions,\cite{Yost.2013} size-consistency,\cite{Yost.2016} and arbitrary determinant expansions.\cite{Yost.2018} In an alternative approach, Nite and Jim\'{e}nez-Hoyos reported correlating a \NOCI{} reference by expanding all basis determinants over single and double excitations, resulting in \NOCISD{}.\cite{Nite.2019} However, none of these approaches considered how the correlations on top of one determinant in the \NOCI{} expansion mapped onto the \DODC{} reference space or onto the external space generated from other determinants, leading to linear dependencies and as a consequence, potential double counting of correlation. As nonorthogonal configurations can be expanded in a common orbital basis using a generalized Thouless transformation, the problems associated with the definition of the \NOCI{} external space are similar to those encountered in orthogonal internally-contracted methods. Therefore, redefining internally contracted orthogonal methods for the development of post-\NOCI{} is likely to provide a suitable approach.

To define the external space in multireference methods, such as fully or partially \icMRCI{},\cite{WernerJCP82_76_3144,WernerJCP88_89_5803} the excited configurations are divided into categories based on the number of electrons remaining in the active orbitals after excitation and the orbital subspaces involved in the excitations. The usual definitions for configurations that retain the number of active orbitals are: internal (active-to-active excitation strings), external (closed-to-virtual excitation strings) and mixed (combinations of active-to-active and closed-to-virtual excitation strings). For configurations that modify the number of active orbitals the subspaces are: semiinternal(+) (composed of active-to-virtual excitations) and semiinternal(-) (composed of closed-to-active excitations).\cite{kutzelniggJCP97_107_432,Lischka:2018da}
In the orthogonal case, each of these configuration types, with the exception of active-to-active excitations, are subspaces of the external space. As excitation operators are applied to the entire reference wavefunction in an \icMRCI{} approach, the generated external space basis is nonorthogonal and must be orthogonalized. However, as the intersection between different subspaces is empty, orthogonal methods can take advantage of the fact that configurations in different subspaces are orthogonal by construction, and so orthogonalization of each subset can be performed separately.
We emphasize that the nonorthogonal nature of the \icMRCI{} external subspaces is not the same as that of \NOCI{}. In \icMRCI{}, nonorthogonality arises from different excitations partially mapping onto the same external configuration, despite the use of a single set of orthogonal orbitals. In contrast, \NOCI{} nonorthogonality is intrinsic to the reference space configurations due to the \DODC{} construction, i.e.\ there are several sets of orthogonal orbitals that are nonorthogonal to each other. After generation of an uncontracted post-\NOCI{} external space, a similar external-space nonorthogonality must be treated in the same way as in \icMRCI{} due to distinct excitations partially mapping onto each other. However, an additional issue that only arises in post-\NOCI{} methods is the presence of internal contamination in the excited configurations. 
In this work, we present an approach for defining the external space built from a nonorthogonal reference, which puts post-\NOCI{} approaches on equal footing with orthogonal internally-contracted multireference approaches and so simplifies the conversion of orthogonal methods to the nonorthogonal domain. As a result of the generality of the approach described, it can be used for, but is not limited to, configuration interaction,\cite{WernerJCP82_76_3144,WernerJCP88_89_5803} perturbation\cite{AnderssonJCP92_96_1218,AngeliJCP01_114_10252} and coupled cluster\cite{Hanauer.2011,Sivalingam.2016,EvangelistaJCP18_149_030901} based methods.

Before proceeding, we first outline the notation used in the remainder of this communication. The set of determinants are labeled $I,J,K\ldots$, while the set of states are labeled $A,B,C\ldots$. We use the notation $a,b,c\ldots$ to describe virtual orbitals and $i,j,k\ldots$ to describe occupied (inactive) orbitals in the original determinant basis. These orbitals are constructed from a set of \AOs{} indexed $\nu,\mu,\tau\ldots$  with overlap $S_{\mu\nu}$. In the common \NO{} basis, the virtual space is indicated by indices $\tilde{a},\tilde{b},\tilde{c}\ldots$, $\tilde{i},\tilde{j},\tilde{k}\ldots$ gives the inactive space indices, and $\tilde{u},\tilde{v},\tilde{w}\ldots$ are indices of active space orbitals. General orbitals are denoted $p,q,r\ldots$ and $\tilde{p},\tilde{q},\tilde{r}\ldots$ in the original and common \MO{} basis respectively. The transformation from the orbital basis of determinant $I$ to the common \NO{} basis is given by the coefficients $\{{}^{I}t_{p}^{\tilde{q}}\}$. The creation operators $\{\tilde{a}_{p}^{\dagger}\}$ and $\{{}^{I}{a}_{p}^{\dagger}\}$ create an electron in the \NO{} or determinant orbitals respectively, while $\{\tilde{a}_{p}\}$ and $\{{}^{I}{a}_{p}\}$ are the respective annihilation operators. $\{{}^{I}C_{p\nu}\}$, $\{{}^{I}\tilde{C}_{p\nu}\}$ and $\{\tilde{C}_{p\nu}\}$ are the coefficients that describe how the \MOs{}, biorthogonal orbitals and \NOs{} are constructed from the \AO{} basis, while $\{D_{IA}\}$ are the \CI{}-expansion coefficients. 

Having described the background and motivation for developing a robust protocol for the construction of the post-\NOCI{} external space, we now examine the challenges posed by a nonorthogonal reference wavefunction in more detail. The first issue we consider is that there is no universal set of occupied \MOs{} in the nonorthogonal reference expansion from which arbitrary excitations uniquely map to the external space. In fact, this is not a problem of nonorthogonal expansions, but of any multiconfigurational reference wavefunction. For nonorthogonal expansions, there is an even greater challenge as the lack of any common orbital space inhibits straightforward identification of the internal space. To illustrate this issue, consider two determinants that are strictly nonorthogonal, such that $\langle J\vert I\rangle\ne0$. These determinants are related by a unitary transformation parameterized in terms of single-excitation operator amplitudes ${}^{I}k_{ia}$.\cite{Thouless.1960}
\begin{equation}
    \vert J\rangle = \exp\left(\sum_{ia}{}^{I}k_{ia}{}^{I}a_{a}^{\dagger}{}^{I}a_{i}\right)\vert I\rangle = \left(1+\sum_{ia}{{}^{I}k_{ia}}{{}^{I}a_{a}^{\dagger}}{}^{I}a_{i}+\frac{1}{2}\sum_{ijab}{{}^{I}k_{ia}{}^{I}k_{jb}}{}^{I}a_{a}^{\dagger}{}^{I}a_{i}{}^{I}a_{b}^{\dagger}{}^{I}a_{j}+\ldots\right)\vert I\rangle.\label{eq:Thouless}
\end{equation}
From eq.\ \ref{eq:Thouless} it is apparent that the set of occupied orbitals in $\vert I\rangle$ cannot be easily partitioned into occupied and virtual orbitals in $\vert J\rangle$, but rather every orbital can map to all occupied and virtual orbitals of $\vert J\rangle$.
Moreover, any excitation over $\vert J\rangle$ can also be described as an excitation over a fully internally contracted expansion in terms of orbitals of $\vert I\rangle$.

Eq.\ \ref{eq:Thouless} also illustrates how, unlike for orthogonal methods, the \NOCI{} wavefunction representation may require a full \CI{} expansion when the \NOCI{} reference determinants are transformed to a common set of \MOs{}. To avoid the intractable issues associated with the use of a common basis, Burton and Thom built the first-order wavefunction by performing determinant-specific substitutions over each term in the expansion, i.e.\ excitations in the \MO{} basis of $\vert I\rangle$ are applied only to configuration $\vert I\rangle$.\cite{Burton.2019}
This expansion over the nonorthogonal reference space scales with the number of reference configurations, $\mathcal{O}(N_{ref}N_{elec}^{l}N_{basis}^{l})$ where $l$ is the order of excitation, as for an uncontracted orthogonal approach.\cite{Lischka:2018da,Sivalingam.2016} However, in the nonorthogonal case, as the excitation operators cannot be divided by their action on different orbital subspaces, it is not possible to take advantage of the partitioning of the external space into subspaces. Therefore, the full external space basis must be orthogonalized at once, leading to significantly greater computational expense than the orthogonal case. Furthermore, in addition to generating linear dependencies among the external configurations, the excitations over a nonorthogonal expansion also yield substituted configurations that partially map onto the internal space. For example, while a single occupied-to-virtual excitation in the \MO{} basis of configuration $\vert{I}\rangle$ will never map back onto $\vert{I}\rangle$, it can map onto any of the other reference configurations as
\begin{equation}
|\langle { J}\vert ^{ I} a_a^{\dagger} {^{ I} a_i} \vert{ I}\rangle| \geq 0 \implies ^{ I} a_a^{\dagger} {^{ I} a_i} \vert{ I}\rangle\in \mathcal{H},
\end{equation}
where we define the Hilbert space as $\mathcal{H}=[\{\vert{I}\rangle\}]\oplus [\{\vert ^\perp{ I}\rangle\}]$, in which $[\{\vert ^\perp{ I}\rangle\}]$ is the external space that contains all configurations that are perpendicular to all internal configurations.

Based on Burton and Thom's work,\cite{Burton.2019} the external space is generated by removing the internal contamination from all excitations over the set of internal configurations in their respective \MO{} basis. To remove the internal space component from the excited configurations, we define the projector 
\begin{align}
P=\sum_{IJ}\vert I\rangle\langle I\vert J\rangle^{-1}\langle J\vert=\sum_A |\Psi_A\rangle\langle\Psi_A|,\label{eq:projector_operator}
\end{align}
with $\{|\Psi_A\rangle\}$ being the states of the nonorthogonal expansion. 
The proposed projector differs from that of Burton and Thom, where they remove only the reference wavefunction from the excited space.\cite{Burton.2019} The removal of only the reference wavefunction from the excitations can be interpreted as a reference relaxation strategy for fixed-reference methods.
However, in general, the entire internal space contamination should be removed from the external space. As an example, for a single excitation, an external-space configuration is given by
\begin{align} 
\vert{}^\perp I_{i}^{a}\rangle&= {{}^I a_a^\dagger} {}^I a_i|I\rangle -\sum_{JK} \vert J \rangle \langle J\vert K\rangle^{-1}\langle K|{}^I a_a^\dagger {}^I a_i|I\rangle= |I_{i}^{a}\rangle -\sum_{JK} \vert J \rangle\langle J\vert K\rangle^{-1}\langle K|I_{i}^{a}\rangle.\label{eq:uncontracter_MO_NO}
\end{align}
in which the projector of eq.\ \ref{eq:projector_operator} ensures contamination from the internal space is removed. 

 Once the internal space component is removed from the generated external space, the remaining external space basis will be nonorthogonal and linearly dependent, as is also the case in orthogonal multireference methods.\cite{Hanauer.2011,Sivalingam.2016,Andersson.1990}
The linear dependence of this external space can be removed via canonical orthogonalization, followed by the removal of singular and near-singular components according to
\begin{align} 
\mathbf{U^\dagger {^{\perp}S}U} = \mathbf{s}\label{eq:perpendicular_overlap},
\end{align} 
where $\mathbf{U}$ is a unitary transformation that diagonalizes the external space basis overlap matrix ${^{\perp}\mathbf{S}}$. The canonical transformation matrix can then be defined as:
\begin{align} 
\mathbf{X} = \mathbf{U}\mathbf{s}_{\zeta}^{-\frac{1}{2}}\label{eq:perpendicular_transformation},
\end{align} 
where $\mathbf{s}_{\zeta}$ is the diagonal overlap matrix in which diagonal values smaller than a threshold ($\zeta$) are removed.
As a consequence, the transformation $\mathbf{X}$ is likely to be rectangular.

 To resolve the issues resulting from ill-defined excitation types in nonorthogonal methods just discussed, we propose use of the multireference \onePDM{}. Earlier works used the \onePDM{} to construct the generalized Fock operator in the spin orbital basis to define the zeroth order Hamiltonian $\hat{H}_{0}$.\cite{Andersson.1990,Andersson.1992,Burton.2020} In contrast, our approach uses the \onePDM{} in a conceptually different way -- instead of the spin-orbital basis,  we propose that the \NOs{} obtained from the \onePDM{} can be used as a common \MO{} basis to construct the excitation operators. The \NOs{} generate three sets of orbitals depending on their fractional occupation: collective closed (full occupation), active and virtual (zero occupation). The closed set can be expanded in terms of orbitals that are occupied in all configurations in the expansion, while the virtual set in terms of orbitals that are unoccupied in all configurations. As a result, the method allows a clear division between excitations that only map onto the external space, and those that can map the entire space.

\begin{figure}[h!]
\centering
\input{orbs}
\caption{Depiction of a small nonorthogonal expansion of configurations in their original molecular orbitals basis and the set of natural orbital with the proposed orbital partitioning.
Closed (inactive) natural orbitals map entirely on to orbitals that are occupied in all nonorthogonal determinants, while virtual natural orbitals map entirely on to orbitals that are unoccupied. The fractionally occupied active natural orbitals map on to orbitals that are occupied in some configurations but virtual in other configurations.
}
\label{fig:orbital_partition}
\end{figure}
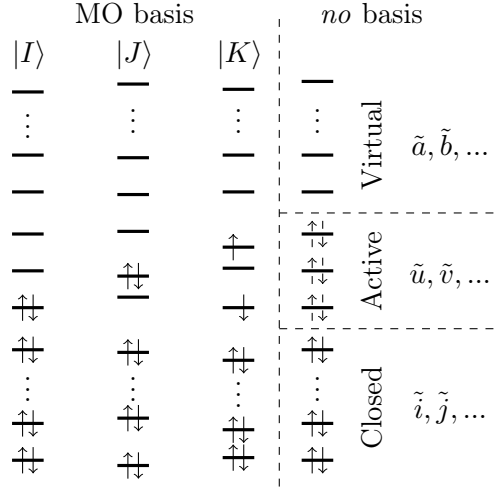

Having outlined the shortcomings of using configuration-specific excitations in an \NOCI{} uncontracted approach, we propose the use of \NOs{} as a unified excitation basis. The set of \NOs{} is divided such that $\tilde B_v = \{{\tilde a_a}:|\langle|{\tilde a_a} {^{I} a_i^\dagger}|\rangle|\le\epsilon, \forall \vert I\rangle\}$, $\tilde B_o = \{{\tilde a_i}:|\langle|{\tilde a_i} {}^{I} a_{a}{^\dagger}|\rangle|\le\epsilon,  \forall \vert I\rangle \}$ and $\tilde B_a = \{{\tilde a_u}:\langle|{\tilde a_u} {\tilde a_i^\dagger}|\rangle = \langle|{\tilde a_u} {\tilde a_a^\dagger}|\rangle=0, \forall \tilde{a}_a\in \tilde{B}_v,\tilde{a}_i\in \tilde{B}_o\}$, where $ dim(\tilde B_o) + dim(\tilde B_a) + dim(\tilde B_v) = N_{basis}$ and $\epsilon$ is a threshold close to zero. Fig.\ \ref{fig:orbital_partition} gives a simplified comparison between a set of nonorthogonal reference configurations in their orthogonal \MO{} basis sets and the proposed \NO{} basis partition.
For the purpose of remainder of this communication, we assume that the \NOs{} are constructed from the total one-particle density matrix. Therefore, any ${\tilde a_a}\in\tilde B_v$ can be expanded using only orbitals that are unoccupied in all reference configurations:
\begin{align} 
{\tilde a_a} = \sum_b {}^{I}a_{b} {}^{I}t_{b}^{\tilde{a*}}\Rightarrow {}^{I}t_{b}^{\tilde {a}}=\langle\vert{}^{I}a_{b}\tilde{a}_{a}^{\dagger}\vert\rangle=\sum_{\mu\nu}{}^{I}C_{b\mu}^{*}S_{\mu\nu}\tilde{C}_{a\nu};\label{eq:no_virtual}
\end{align}
and, similarly, the elements of $\tilde B_o$ are composed of only orbitals that are occupied in all configurations:
\begin{align} 
\tilde{a}_{i} = \sum_j {}^{I}a_{j} {}^{I}t_{j}^{\tilde{i}*}\Rightarrow {^{I} t_j^{\tilde i}}=\langle|{^{I} a_j}{\tilde a_i^\dagger}|\rangle=\sum_{\mu\nu}{}^{I}C_{j\mu}^{*}S_{\mu\nu}\tilde{C}_{i\nu}.\label{eq:no_closed}
\end{align}
Based on these definitions, $\tilde B_v$ spans the total virtual space of the \NO{} expansion and $\tilde B_o$ spans the total occupied space.
The remaining orbitals compose the elements of an active space $\tilde B_a$, which requires the full \MO{} orbital space of at least one reference configuration to be represented:
\begin{align} 
{\tilde a_u} = \sum_p {^{I} a_p {^{I} t_p^{\tilde u*}}}\Rightarrow {^{I} t_p^{\tilde u}}=\langle|{^{I} a_p}{\tilde a_u^\dagger}|\rangle=\sum_{\mu\nu}{}^{I}C_{p\mu}^{*}S_{\mu\nu}\tilde{C}_{u\nu}.\label{eq:no_active}
\end{align}

In the \NO{}-basis representation, the internal contamination in the space spanned by excitations is restricted to excitation operator strings comprised exclusively of $\tilde B_a$ and $\tilde B_a^{\dagger}$. 
Based on Burton's unified Wick's theorem,\cite{BurtonJCP21_154_144109} we can show the behavior of each \NO{} subspace via their symmetrical contraction with respect to the asymmetric Fermi vacuum $\langle I\vert ... \vert J\rangle$. 
Only one type of contraction containing virtual \NOs{} is non-zero due to the definition of $\tilde B_v$:
\begin{align}
 \wick{\c1{\tilde a}_{a} \c1{\tilde a}_{p}^\dagger}=\wick{\c1{\tilde a}_{p} \c1{\tilde a}_{a}^\dagger}&=\delta_{0,\text{dim}({}^{IJ}\mathcal{N})}\delta_{\tilde a\tilde p},\label{eq:wick_ap}
\end{align} 
where $\delta_{ij}$ is Kronecker delta and ${}^{IJ}\mathcal{N}$ is the null space between occupied orbitals of $ \vert I\rangle$ and $ \vert J\rangle$.
In other words, $\text{dim}({}^{IJ}\mathcal{N})$ is the number of occupied orbitals with zero-overlap in the biorthogonal basis between $ \vert I\rangle$ and $ \vert J\rangle$ configurations.
Similarly, for contractions involving closed \NOs{}, it can be shown that due to the definition of $\tilde B_o$ the only non-zero contractions are:
\begin{align}
\wick{\c1{\tilde a}_{j}^\dagger \c1{\tilde a}_{i}}=& \sum_{\{k|{}^{IJ}\tilde{N}_{k}\neq 0\}}\Bigg(\sum_{\nu\mu}{\tilde C_{j\nu}^*}{S_{\nu\mu}}{}^I{\tilde C_{k\mu}}\Bigg){}^{IJ}\tilde{N}_{k}^{-1}\Bigg(\sum_{\tau\chi}{}^J{\tilde C^*_{k\tau}} {S_{\tau\chi}}{\tilde C_{i\chi}}\Bigg)\delta_{0,\text{dim}({}^{IJ}\mathcal{N})} \nonumber \\
&+\sum_{\{k|{}^{IJ}\tilde{N}_{k}= 0\}}\Bigg(\sum_{\nu\mu}{\tilde C_{j\nu}^*}{S_{\nu\mu}}{}^I{\tilde C_{k\mu}}\Bigg)\Bigg(\sum_{\tau\chi}{}^J{\tilde C^*_{k\tau}} {S_{\tau\chi}}{\tilde C_{i\chi}}\Bigg)(\delta_{0,\text{dim}({}^{IJ}\mathcal{N})}+\delta_{1,\text{dim}({}^{IJ}\mathcal{N})}) \nonumber \\
&+\sum_{\{k|{}^{IJ}\tilde{N}_{k}= 0\}}\Bigg(\sum_{\nu\mu}{\tilde C_{j\nu}^*}{S_{\nu\mu}}{}^J{\tilde C_{k\mu}}\Bigg)\Bigg(\sum_{\tau\chi}{}^J{\tilde C^*_{k\tau}} {S_{\tau\chi}}{\tilde C_{i\chi}}\Bigg)\delta_{0,\text{dim}({}^{IJ}\mathcal{N})}\label{eq:wick_pi}\\
\wick{\c1{\tilde a}_{i} \c1{\tilde a}_{j}^\dagger}&=\delta_{0,\text{dim}({}^{IJ}\mathcal{N})}\delta_{{\tilde i}{\tilde j}} - \wick{\c1{\tilde a}_{j}^\dagger \c1{\tilde a}_{i}},\label{eq:wick_ip}
\end{align}   
where ${}^J{\tilde C_{k\tau}}$ are the \MO{} coefficients in the biorthogonal basis between $ \vert I\rangle$ and $ \vert J\rangle$, and ${}^{IJ}\tilde{N}_{k}$ is the orbital overlap in the same biorthogonal basis.
Lastly, the active-active contractions behaves as a symmetrical nonorthogonal contraction,\cite{BurtonJCP21_154_144109}
\begin{align}
\wick{\c1{\tilde a}_{u}^\dagger \c1{\tilde a}_{v}}=& \sum_{\{j|{}^{IJ}\tilde{N}_{j}\neq 0\}}\Bigg(\sum_{\nu\mu}{\tilde C_{u\nu}^*}{S_{\nu\mu}}{}^I{\tilde C_{j\mu}}\Bigg){}^{IJ}\tilde{N}_{j}^{-1}\Bigg(\sum_{\tau\chi}{}^J{\tilde C^*_{j\tau}} {S_{\tau\chi}}{\tilde C_{v\chi}}\Bigg)\delta_{0,\text{dim}({}^{IJ}\mathcal{N})} \nonumber \\
&+\sum_{\{j|{}^{IJ}\tilde{N}_{j}= 0\}}\Bigg(\sum_{\nu\mu}{\tilde C_{u\nu}^*}{S_{\nu\mu}}{}^I{\tilde C_{j\mu}}\Bigg)\Bigg(\sum_{\tau\chi}{}^J{\tilde C^*_{j\tau}} {S_{\tau\chi}}{\tilde C_{v\chi}}\Bigg)(\delta_{0,\text{dim}({}^{IJ}\mathcal{N})}+\delta_{1,\text{dim}({}^{IJ}\mathcal{N})}) \nonumber \\
&+\sum_{\{j|{}^{IJ}\tilde{N}_{j}= 0\}}\Bigg(\sum_{\nu\mu}{\tilde C_{u\nu}^*}{S_{\nu\mu}}{}^J{\tilde C_{j\mu}}\Bigg)\Bigg(\sum_{\tau\chi}{}^J{\tilde C^*_{j\tau}} {S_{\tau\chi}}{\tilde C_{v\chi}}\Bigg)\delta_{0,\text{dim}({}^{IJ}\mathcal{N})},\label{eq:wick_uv}\\
\wick{\c1{\tilde a}_{u} \c1{\tilde a}_{v}^\dagger}&=\delta_{0,\text{dim}({}^{IJ}\mathcal{N})}\delta_{{\tilde u}{\tilde v}} - \wick{\c1{\tilde a}_{v}^\dagger \c1{\tilde a}_{u}}.\label{eq:wick_vu}
\end{align}   

Using the \NO{} contractions and the excitation determinant notation,\cite{BurtonJCP21_154_144109,BurtonJCP22_157_204109} we can show that applying an arbitrary string of excitation operators containing either a closed \NO{} annihilation operator or a virtual \NO{} creation operator over any reference configuration has a zero projection onto the internal space,                            
\begin{align}
\langle I\vert{\tilde a_{a}^\dagger{\tilde a_{r}}\bigg({\prod_{n=1}^l\tilde a_{p_n}^\dagger}{\tilde a_{q_n}}\bigg)}\vert J\rangle ={}^{IJ}{\tilde N}
\begin{vmatrix}
     \wick{\c1{\tilde a}_{a}^\dagger \c1{\tilde a}_{r}} & \wick{\c1{\tilde a}_{r} \c1{\tilde a}_{p_1}^\dagger} & ... &\wick{\c1{\tilde a}_{r} \c1{\tilde a}_{p_l}^\dagger}\\ 
     \wick{\c1{\tilde a}_{a}^\dagger \c1{\tilde a}_{q_1}} & \wick{\c1{\tilde a}_{p_1}^\dagger \c1{\tilde a}_{q_1}} & ...& \wick{\c1{\tilde a}_{q_1} \c1{\tilde a}_{p_l}^\dagger}\\ 
     \vdots & \vdots & \ddots &\vdots \\
      \wick{\c1{\tilde a}_{a}^\dagger \c1{\tilde a}_{q_l}} & \wick{\c1{\tilde a}_{p_1}^\dagger \c1{\tilde a}_{q_l}} & ... & \wick{\c1{\tilde a}_{p_l}^\dagger \c1{\tilde a}_{q_l}}
\end{vmatrix}
={}^{IJ}{\tilde N}
\begin{vmatrix}
     0 & \wick{\c1{\tilde a}_{r} \c1{\tilde a}_{p_1}^\dagger} & ... &\wick{\c1{\tilde a}_{r} \c1{\tilde a}_{p_l}^\dagger}\\ 
     0  & \wick{\c1{\tilde a}_{p_1}^\dagger \c1{\tilde a}_{q_1}} & ...& \wick{\c1{\tilde a}_{q_1} \c1{\tilde a}_{p_l}^\dagger}\\ 
     \vdots & \vdots & \ddots &\vdots \\
      0 & \wick{\c1{\tilde a}_{p_1}^\dagger \c1{\tilde a}_{q_l}} & ... & \wick{\c1{\tilde a}_{p_l}^\dagger \c1{\tilde a}_{q_l}}
\end{vmatrix}=0, \forall q_n,r\notin \tilde B_v,\label{eq:overlap_no_virtual}
\end{align}    
\begin{align}
\langle I\vert{\tilde a_{r}^\dagger{\tilde a_{i}}\bigg({\prod_{n=1}^l\tilde a_{p_n}^\dagger}{\tilde a_{q_n}}\bigg)}\vert J\rangle ={}^{IJ}{\tilde N}
\begin{vmatrix}
     \wick{\c1{\tilde a}_{r}^\dagger \c1{\tilde a}_{i}} & \wick{\c1{\tilde a}_{i} \c1{\tilde a}_{p_1}^\dagger} & ... &\wick{\c1{\tilde a}_{i} \c1{\tilde a}_{p_l}^\dagger}\\ 
     \wick{\c1{\tilde a}_{r}^\dagger \c1{\tilde a}_{q_1}} & \wick{\c1{\tilde a}_{p_1}^\dagger \c1{\tilde a}_{q_1}} & ...& \wick{\c1{\tilde a}_{q_1} \c1{\tilde a}_{p_l}^\dagger}\\ 
     \vdots & \vdots & \ddots &\vdots \\
      \wick{\c1{\tilde a}_{r}^\dagger \c1{\tilde a}_{q_l}} & \wick{\c1{\tilde a}_{p_1}^\dagger \c1{\tilde a}_{q_l}} & ... & \wick{\c1{\tilde a}_{p_l}^\dagger \c1{\tilde a}_{q_l}}
\end{vmatrix}
={}^{IJ}{\tilde N}
\begin{vmatrix}
     0 & 0 & ... &0\\ 
     \wick{\c1{\tilde a}_{r}^\dagger \c1{\tilde a}_{q_1}} & \wick{\c1{\tilde a}_{p_1}^\dagger \c1{\tilde a}_{q_1}} & ...& \wick{\c1{\tilde a}_{q_1} \c1{\tilde a}_{p_l}^\dagger}\\ 
     \vdots & \vdots & \ddots &\vdots \\
      \wick{\c1{\tilde a}_{r}^\dagger \c1{\tilde a}_{q_l}} & \wick{\c1{\tilde a}_{p_1}^\dagger \c1{\tilde a}_{q_l}} & ... & \wick{\c1{\tilde a}_{p_l}^\dagger \c1{\tilde a}_{q_l}}
\end{vmatrix}=0, \forall p_n,r\notin \tilde B_o,\label{eq:overlap_no_closed}
\end{align}    
with ${}^{IJ}{\tilde{N}}=\prod_{\{j|{}^{IJ}{\tilde{N_j}\neq 0\}}}{}^{IJ}{\tilde{N_j}}$ being the pseudo overlap.

Equations~\eqref{eq:overlap_no_virtual} and \eqref{eq:overlap_no_closed} have two important consequences.
First, in a nonorthogonal \NO{}-excitation based framework, internal contamination can only occur in excitation strings composed exclusively of active orbitals. 
Therefore, active-to-active excitations are the only set that cannot be easily assigned as belonging to the internal or external space, and so must be purified of internal contamination. 
As in eq.~\eqref{eq:uncontracter_MO_NO}, the component of an active-to-active substitution that belongs to the external space is
\begin{align} 
|{^{\perp}{{I}_{{\tilde u}_1...{\tilde u}_l}^{{\tilde v}_1...{\tilde v}_l}}}\rangle&=(1-P)|{{I_{{\tilde u}_1...{\tilde u}_l}^{{\tilde v}_1...{\tilde v}_l}}}\rangle = \bigg({\prod_{n=1}^l\tilde a_{v_n}^\dagger}{\tilde a_{u_n}}\bigg)|I\rangle-\sum_{JK}|J\rangle\langle J\vert K\rangle^ {-1}\langle K|\bigg({\prod_{n=1}^l\tilde a_{v_n}^\dagger}{\tilde a_{u_n}}\bigg)|I\rangle.
\label{eq:uncontracted_no_NO}
\end{align}
Second, the general string of excitations can be divided at an arbitrary point into bra and a ket excitations.
As a result of the contraction relationships, the overlap between the resulting two excited configurations is zero if the number of creation and annihilation operators in each of the bra and ket excitation strings differ in the number of \NO{} virtual or \NO{} closed orbitals.
Therefore, the excited space of a string of \NO{} excitations can be divided in subspaces as
\begin{align}
[\{\vert I_{\tilde{p}_1...\tilde{p}_l}^{\tilde{q}_1...\tilde{q}_l}\rangle\}] =\bigoplus_{v,o=0}^l  [\{\vert  I^{(l,o,v)}\rangle\}],\label{eq:space_division}
\end{align}
where $\{\vert I^{(l,o,v)}\rangle\}$ is the set of $l$-order \NO{} substituted determinants with $o$ \NO{} closed orbital and $l-o$ \NO{} active orbital annihilation operators, and $v$ \NO{} virtual orbital and $l-v$ \NO{} active orbital creation operators.
Therefore, it is possible to partition the external space configurations depending on the orbital subspace on which excitation operators act, avoiding the requirement to construct the full external space basis overlap matrix to remove linear dependencies. Furthermore, we reinforce that only the $(l,0,0)$ block comprising of only \NO{} active orbital annihilation/creation operator strings contains internal space contamination which must be removed through eq.\ \ref{eq:uncontracted_no_NO}. As a result, the excited space shown in eq.\ \ref{eq:space_division} can be reduced to the external space as:
\begin{align}
[\{\vert ^\perp I_{\tilde{p}_1...\tilde{p}_l}^{\tilde{q}_1...\tilde{q}_l}\rangle\}] =\Bigg(\bigoplus_{\substack{v,o=0 \\ v+o>0}}^l  [\{\vert  I^{(l,o,v)}\rangle\}]\Bigg)\oplus [\{\vert ^\perp I^{(l,0,0)}\rangle\}].\label{eq:external_space_division}
\end{align}

We now demonstrate that the external/excited spaces generated by excitations up to order $l$ in the natural excitation basis span the same space as \MO{} excitations up to the same order on top of every reference determinant. I.e.\ any $l$-order excitation constructed in the \NO{} basis can be expressed as a linear combination of up to $l$-order \MO{} excitations on all reference determinants (and \textit{vice versa}). Consequently, there is no loss of accuracy of when using an external space constructed using a natural excitation basis up to a given order (assuming a sufficiently small $\epsilon$ threshold during the \NO{} subspace definition).
First, using eqs.\ \ref{eq:no_virtual}, \ref{eq:no_closed} and \ref{eq:no_active}, it can be shown that any string of $l$ \NO{} excitations over a reference configuration is contained in the space spanned by \MO{} excited configurations up to $l$-order, 
\begin{align}
    \vert {I}_{\tilde{p}_1...\tilde{p}_l}^{\tilde{q}_1...\tilde{q}_l}\rangle =  \bigg({\prod_{n=1}^l\tilde{a}_{q_n}^\dagger}{\tilde{a}_{p_n}}\bigg)\vert I\rangle = {\prod_{n=1}^l}\bigg(\sum_{r_n s_n} {}^It_{s_n}^{\tilde{p}_n*}{}^It_{r_n}^{\tilde{q}_n} {{}^I{a}_{r_n}^\dagger}{{}^I{a}_{s_n}}\bigg)\vert I\rangle =\sum_{\substack{s_1...s_l\\ r_1...r_l}} {\prod_{n=1}^l}\bigg( {}^It_{s_n}^{\tilde{p}_n*}{}^It_{r_n}^{\tilde{q}_n}\bigg)\vert I_{s_1...s_l}^{r_1...r_l}\rangle.  
\end{align}
Therefore, in general 
$[\{\vert  {I}_{\tilde{p}_1...\tilde{p}_l}^{\tilde{q}_1...\tilde{q}_l}\rangle\}]\subset [\{\vert I_{s_1...s_l}^{r_1...r_l}\rangle\}] = [\{\vert {I}\rangle\}] + [\{\vert {I}_{i}^{a}\rangle\}]+...+[\{\vert {I}_{i_1...i_l}^{{a}_1...{a}_l}\rangle\}]$ for the excited space, while for the external space, in which internal contamination has been removed from the excited space, $[\{\vert ^\perp {I}_{\tilde{p}_1...\tilde{p}_l}^{\tilde{q}_1...\tilde{q}_l}\rangle\}]\subset [\{\vert ^\perp I_{s_1...s_l}^{r_1...r_l}\rangle\}] = [\{\vert ^\perp{I}_{i}^{a}\rangle\}]+...+[\{\vert ^\perp{I}_{i_1...i_l}^{{a}_1...{a}_l}\rangle\}]$.
In the specific case where only closed to virtual \NO{} excitations are considered, $[\{\vert {I}_{\tilde{i}_1...\tilde{i}_l}^{\tilde{a}_1...\tilde{a}_l}\rangle\}]\subset [\{\vert I_{i_1...i_l}^{a_1...a_l}\rangle\}]$, where the sets can only be equal if the reference is a single determinant.
In a similar way, any string of \MO{} excitations over a reference can be expanded in terms of \NO{} excitations,
\begin{align}
    \vert {I}_{i_1...i_l}^{a_1...a_l}\rangle = \bigg({\prod_{n=1}^l{{}^Ia}_{a_n}^\dagger}{{{}^Ia}_{i_n}}\bigg)\vert I\rangle = {\prod_{n=1}^l\bigg(\sum_{\tilde{p}_n \tilde{q}_n} {}^It_{i_n}^{\tilde{p}_n}{}^It_{a_n}^{\tilde{q}_n*}\tilde{a}_{q_n}^\dagger}{\tilde{a}_{p_n}}\bigg)\vert I\rangle= \sum_{\substack{\tilde{p}_1...\tilde{p}_l\\ \tilde{q}_1...\tilde{q}_l}}{\prod_{n=1}^l\bigg( {}^It_{i_n}^{\tilde{p}_n}{}^It_{a_n}^{\tilde{q}_n*}}\bigg)\vert I_{\tilde{p}_1...\tilde{p}_l}^{\tilde{q}_1...\tilde{q}_l}\rangle,
\end{align}  
which implies that $ [\{\vert I_{i_1...i_l}^{a_1...a_l}\rangle\}]\subset[\{\vert {I}_{\tilde{p}_1...\tilde{p}_l}^{\tilde{q}_1...\tilde{q}_l}\rangle\}]$ and, more generally, that $  [\{\vert I_{r_1...r_l}^{s_1...s_l}\rangle\}]\subset[\{\vert {I}_{\tilde{p}_1...\tilde{p}_l}^{\tilde{q}_1...\tilde{q}_l}\rangle\}]$. As $\tilde{p}_{n}$ and $\tilde{q}_{n}$ run over all \NOs{}, the excitations include redundant terms, e.g.\ $\tilde{a}_{i}^{\dagger}\tilde{a}_{i}$ or $\tilde{a}_{a}\tilde{a}_{a}^{\dagger}$. This situation is similar to internally-contracted multireference methods, where $l$-order mixed and semiinternal excitations generate a set of excited determinants up to order $l$.\cite{Hanauer.2011} In addition, unless a \CAS{} is used, the set of $l$-order excitations does not necessarily generate the complete lower-order excited space, as is the case in post-\NOCI{} methods.   
As a consequence, the partition of $[\{\vert {I}_{\tilde{p}_1...\tilde{p}_l}^{\tilde{q}_1...\tilde{q}_l}\rangle\}]$ in eq.~\eqref{eq:space_division} and eq.\ \ref{eq:external_space_division} must account for lower-order excitations, such that the set of intersecting subspaces generated by specific \MO{} excitations can be represented by the direct sum of independent \NO{} excitations as 
\begin{align}
\bigoplus_{v,o=0}^l \Bigg(\sum_{\substack{m=0\\m\ge max(v,o)}}^l [\{\vert I^{(m,o,v)}\rangle\}]\Bigg) = [\{\vert {I}\rangle\}] + [\{\vert {I}_{i}^{a}\rangle\}]+...+[\{\vert {I}_{i_1...i_l}^{{a}_1...{a}_l}\rangle\}] \label{eq:excited_mo_no}
\end{align}
for the excited space, and 
\begin{align}
\bigoplus_{v,o=0}^l \Bigg(\sum_{\substack{m=1\\m\ge max(v,o)}}^l [\{\vert ^\perp I^{(m,o,v)}\rangle\}]\Bigg) = [\{\vert ^\perp{I}_{i}^{a}\rangle\}]+...+[\{\vert ^\perp{I}_{i_1...i_l}^{{a}_1...{a}_l}\rangle\}] \label{eq:external_mo_no}
\end{align}
for the external space. The results of eqs.\ \ref{eq:excited_mo_no} and \ref{eq:external_mo_no} demonstrate that the \NO{} excitation framework does not alter the external space, but allows the space to be partitioned in a more computationally useful manner. 

\begin{figure}[h!]
\centering
\includegraphics{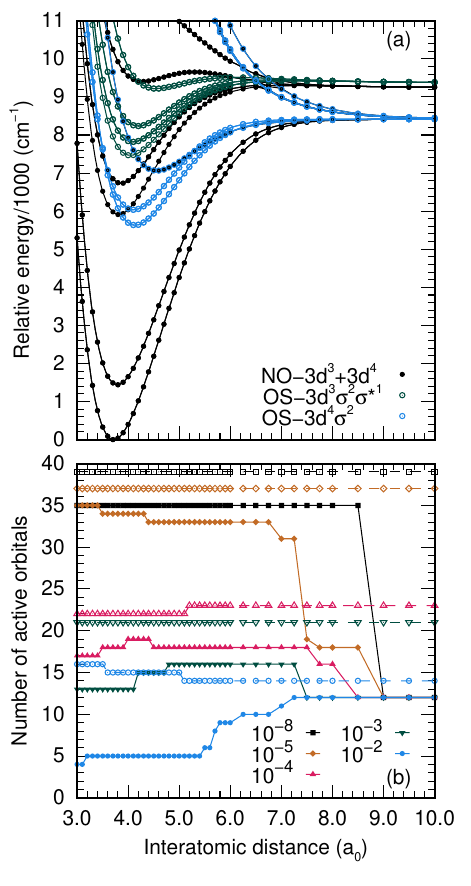}
\caption{
(a) Quintet and septet vanadium monohydride (VH) dissociation potential energy curves using a valence double zeta-type basis set.
Hollow marks indicates the reference occupation-specific (OS) states and filled marks are the states formed by the nonorthogonal coupled OS reference states.
(b) Number of orbitals with natural occupation ($n_o$) between zero and one given a color and shaped coded threshold ($\nu<$ $n_o$ $<1-\nu$).
Dashed lines and hollow markers represents the occupation of the full 1PDM, while full lines and markers represents the the state-specific (ground state 1PDM) occupations.
As a consequence, for a given occupation threshold, the dashed lines are a ceiling for the state-specific spaces.
}
\label{fig:VH}
\end{figure}

To provide a numerical example that demonstrates the advantages expected with the use of the proposed natural excitation framework, we consider the dissociation of vanadium monohydride (VH).
In VH, the $^5\Delta$ ground state has a large mixing between $3d^3$ and $3d^4$-configurations, for which the optimal orbitals are significantly different energetically.
To account for the orbital relaxation effects between configurations, active-space-based orthogonal approaches require inclusion of an additional $d$-shell in the active space (the double-$d$-shell effect),\cite{Moraes.2023} while nonorthogonal methods can use optimized sets of $d$-orbitals for each relevant configuration.
The nonorthogonal wavefunction expansion is constructed from two interacting \OS{} \RASSCF{} expansions, which were independently optimized for $3d^3$ and $3d^4$-occupations using the partially fixed reference space protocol.\cite{moraesPCCP24_26_19742}
Fig.~\ref{fig:VH}(a) depicts the lowest-lying quintet and septet nonorthogonal states, in which mixed configuration character is permitted, compared with the underlying \OS{} states. 
The substantial reduction in energy of the ground and first excited state ($^5\Delta$ and $^5\Pi$, respectively) show the mixed-occupation nature of VH. 
The overlap between the $3d$-orbitals optimized for $d^3$ and $d^4$-configurations is larger than 0.9, confirming these orbitals are qualitative the same.
On the other hand, constructing a $3d^4$ configuration using $3d^3$-optimal \MOs{} results in an energy increase of 0.01 to 0.1 Hartree, depending on the specific configuration, when compared with use of the  $3d^4$ optimized \MOs{}, and \textit{vise-versa}.
Therefore, the improved description of the low-energy states through the use of a nonorthogonal expansion justifies the use of VH as a model system.

Fig.~\ref{fig:VH}(b) shows $dim(\tilde B_a)$ for different occupation number thresholds ($\nu$).  Of note, the \NO{} occupation number is closely related, but not equivalent to the $\epsilon$ used to defined \NO{} subspaces, although in practice the occupation numbers can be used.
Using the full \onePDM{} combined with the most strict occupation number threshold ($\nu =10^{-8}$, which for this case translates in a $\epsilon$ of around 10$^{-6}$) yields $dim(\tilde B_a)=39$ at all geometries along the potential energy curve, out of 96 total \NOs{}.
This space is composed of vanadium $3s$, $3p$, $3d$, $4d$, $4s$ and $5s$ alpha and beta orbitals, vanadium $6s$ and $6p$ alpha orbitals, hydrogen $1s$ alpha and beta orbitals, and a hydrogen $2s$ alpha orbital.
Therefore, the small 90 determinant nonorthogonal expansion is found to include the double-$d$-shell effect.
This relatively large set of active orbitals is a worst case scenario, generated by the amount of near-degenerate open-shell orbitals and small number of localized bonding orbitals.
As a consequence, the number of spin-conserving active-to-active single excitations (773) is slightly smaller than in the original \MO{} basis (864), which reduces the terms from which internal contamination must be removed (as discussed in eq.~\ref{eq:uncontracted_no_NO}).
A similar sized and two smaller \NOs{} single excitation blocks generated by active-to-virtual (904), occupied-to-active (195) and occupied-to-virtual (235), respectively, must also to be considered (scalings are summarized in Table~\ref{tab:S_matrix}).
As a result, for an uncontracted approach, the \NO{} representation can either increase or decrease the total number of external configurations depending on the system, size of expansion and orbital optimization protocol used.

A larger number of redundant external configurations in the \NOs{} excitation basis does not directly imply an increase in computational cost or scaling.
For example, provided the reference wavefunction is qualitatively correct, the number of external configurations with internal contamination generated by excitations in the original \MO{} basis scales with the virtual space and so depends on the basis set size, while the number generated by \NO{} excitations only depends on the $dim(\tilde B_a)$ which remains constant or near-constant.
However, the main computational cost results from construction of the Hamiltonian matrix. For this application, the \NO{}-based framework also has a computational advantage compared with the \MO{} approach. As proved by Burton using the generalized Wick's theorem,\cite{BurtonJCP21_154_144109} any one and two-body matrix element between two excited nonorthogonal configurations can be computed through use of the screened overlap terms and the \MO{} transformed integrals. In the \MO{} excitation approach, the former requires $N_{ref}^2$ square matrices containing $(N_{elec}N_{virtual})^2$ elements and, the latter requires $N_{ref}$ basis transformations per integral. In the \NO{} excitation basis, the number of integral transformations does not scale with the number of reference configurations, as only the \NO{} representation is required. As a result, the number of unique screened overlap terms is reduced to $N_{ref}(N_{ref}/2 +1)$ due to excitation symmetry. Additionally, as a consequence of the definitions of $\tilde{B}_v$ and $\tilde{B}_o$, these matrices have a predefined block diagonal structure, similar to equations~\eqref{eq:wick_ap}-\eqref{eq:wick_vu}.
Although, method-specific optimizations are possible to further reduce the computational cost, these are beyond the scope of this communication, one of the goals of which is to encourage development of these methods.

Uncontracted methodologies have many advantageous properties, nevertheless their scaling with the reference space dimension are prohibitive for the majority of systems.
As such, internally contracted methods are much more predominant, trading scaling for other computational issues, e.g.\ recovering less correlation than their uncontracted counterpart, difficulties properly modeling avoided crossing regions, and more complex calculation of Hamiltonian matrix elements.\cite{Lischka:2018da}
 Such an approach would be inefficient in the \MO{}-based excitation framework, as either the \MO{} representation of one configuration must be chosen to detriment of the other configurations, or all excitations must be applied to all configurations in the \MO{} representation. In the former case, either a configuration is treated differently from the others and so the errors associated with the internally contracted approach are not evenly distributed over all configurations or all possible excitations must be considered, scaling as $\mathcal{O}(N_{basis}^{2l})$ for an $l$-th order excitation. In the latter, the number of excitations scales with the number of reference configurations, removing any computational advantage compared to an uncontracted approach. This realization reinforces the advantages of the unified natural excitation framework in a post-\NOCI{} context. In a \NO{} fully internally contracted post-\NOCI{} approach, an arbitrary active-to-active excitation is given by
\begin{align} 
|{^{\perp}{{\Psi_{A}}_{{\tilde u}_1...{\tilde u}_l}^{{\tilde v}_1...{\tilde v}_l}}}\rangle&=(1-P)|{{{\Psi_{A}}_{{\tilde u}_1...{\tilde u}_l}^{{\tilde v}_1...{\tilde v}_l}}}\rangle = \bigg({\prod_{n=1}^l\tilde a_{v_n}^\dagger}{\tilde a_{u_n}}\bigg)|\Psi_A\rangle-\sum_B|\Psi_B\rangle\langle\Psi_B|\bigg({\prod_{n=1}^l\tilde a_{v_n}^\dagger}{\tilde a_{u_n}}\bigg)|\Psi_A\rangle,
\label{eq:external_component_states}
\end{align}
which can be rewritten in terms of configurations as
\begin{align} 
|{^{\perp}{{\Psi_{A}}_{{\tilde u}_1...{\tilde u}_l}^{{\tilde v}_1...{\tilde v}_l}}}\rangle&= \sum_I\Bigg[\bigg({\prod_{n=1}^l\tilde a_{v_n}^\dagger}{\tilde a_{u_n}}\bigg)|I\rangle-\sum_{JK}|J\rangle\langle J\vert K\rangle^ {-1}\langle K|\bigg({\prod_{n=1}^l\tilde a_{v_n}^\dagger}{\tilde a_{u_n}}\bigg)|I\rangle\Bigg] D_{IA} =\sum_I |{^{\perp}{{I}_{{\tilde u}_1...{\tilde u}_l}^{{\tilde v}_1...{\tilde v}_l}}}\rangle D_{IA}.
\label{eq:external_component_config}
\end{align}
This contraction does not reduce the number of intermediate matrices required, but reduces the size of the final Hamiltonian and external space dependence with respect to the reference space dimension. 

For a state-specific calculation, the size of the \NO{} active space can be reduced by considering the contraction during the definition of \NO{} spaces. The definition of the \NO{} basis and the basis partition can be modified by the use of a state-specific \onePDM{}, such that the \NO{} spaces are given by:
$\tilde B'_v = \{{\tilde a'_a}:|\langle|{\tilde a'_a} {^{I} a_i^\dagger}|\rangle|\le \epsilon, \forall \vert I\rangle, |D_{IA}|> \epsilon\}$, $\tilde B'_o = \{{\tilde a'_i}:|\langle|{\tilde a'_i} {{}^{I} a_{a}^\dagger}|\rangle|\leq \epsilon,  \forall \vert I\rangle, |D_{IA}|> \epsilon \}$ and $\tilde B'_a = \{{\tilde a'_u}:\langle|{\tilde a'_u} {{\tilde a'}{}_i^\dagger}|\rangle = \langle|{\tilde a'_u} {{\tilde{a}'}{}_a^\dagger}|\rangle = 0 , \forall \tilde{a}'_a\in \tilde{B}'_v,\tilde{a}'_i\in \tilde{B}'_o \}$.
In this alternative definition $dim(\tilde B_a)\ge dim(\tilde B'_a)$ and, as a result, the number of active-to-active excitations, the most costly class, is likely to be reduced.
As depicted in Fig.~\ref{fig:VH}(b), for a given threshold the state-specific \NO{} active space dimension is significantly smaller than using the full \onePDM{}, which is an upper limit.
In addition, the variation in active space size with interatomic distance is well behaved considering it can only assume integer values.
For $\nu = 10^{-8}$ the only discontinuity is caused by the change from a $3d^3$/$3d^4$ mixed state to pure $3d^4$ state once the dissociation limit is reached. Lower thresholds will suffer from further error due to the removal of near-threshold active orbitals.
This result indicates that the change in active space size should not lead to large discontinuities, other than those caused by the contraction itself, such as those resulting from drastic changes in the wavefunction, the need of reference relaxation or additional excitations over multiple reference states.\cite{bauschlicherCPL17_683_62}

\begin{table}[]
   \caption{Comparison of scaling of terms required for evaluation of an $l^{\text{th}}$-order excitation between molecular-orbital-based uncontracted, natural-orbital-based uncontracted, and natural-orbital-based fully internally contracted approaches.}
    \centering
    \begin{tabular}{c|c|c}
        & \multicolumn{2}{c}{Dimension of external space terms}\\
    Method              & Nonorthogonal  & Internally contaminated \\
    \hline\hline
    \MO{} uncontracted  & $(N_v^lN_o^lN_{ref})^2$ &  $N_v^lN_o^lN_{ref}$\\
    \NO{} uncontracted  & $\sum_{i=0}^l\sum_{m=0}^l({\tilde N}_v^{m}{\tilde N}_a^{2(l-m-i)}{\tilde N}_o^{i}N_{ref})^2$ & ${\tilde N}_a^{2l}N_{ref}$\\
    \NO{} FIC  & $\sum_{i=0}^l\sum_{m=0}^l({\tilde N}_v^{'m}{\tilde N}_a^{'2(l-m-i)}{\tilde N}_o^{'i})^2$ & ${\tilde N}_a^{'2l}$\\
    \end{tabular}
    \label{tab:S_matrix}
\end{table}

In this work, we have established that the natural excitation basis 1) allows configurations to be unambiguously assigned as belonging to the internal or external space, or as requiring purification of internal contamination, 2) minimizes the excitation space requiring internal purification, 3) permits partitioning of the external space to reduce the computational cost associated with orthogonalization, and 4) up to a given order spans the same space as excitations upon each reference determinant.
Calculation of the VH dissociation curve is presented to demonstrate the expected advantages of the proposed approach.
We have intentionally presented the work in a method-agnostic framework for two reasons. First, the natural excitation basis transformation can be shown to provide substantial computational cost reductions while preserving accuracy without any method-specific considerations. Second, our objective is to highlight the similarities between post-\NOCI{} and orthogonal internally contracted multireference methods, allowing facile conversion of pre-established and successful orthogonal approaches to accelerate the development of more quantitative \DODC{} methods.

\section{Acknowledgements}
This work was supported by the U.S. Department of Energy,Office of Science, Basic Energy Sciences, in the Computational and Theoretical Program (Grant No. DE-SC0024507). 

\section{Data Availability}
Data sharing is not applicable to this article as no new data were created or analyzed in this study.

\bibliography{icmrnoci}

\end{document}

%% file: orbital_tickz_def.tex
\newcommand{\orbEmpty}[2]
{
  \begin{scope}
    \draw[very thick] ({#1-0.3},#2) to ({#1+0.3},#2);
  \end{scope}
}
\newcommand{\orbUp}[2]
{
  \begin{scope}
    \draw[very thick] ({#1-0.3},#2) to ({#1+0.3},#2);
    \draw[->] (#1-0.1,#2-0.25) to (#1-0.1,#2+0.25);
  \end{scope}
}
\newcommand{\orbDown}[2]
{
  \begin{scope}
    \draw[very thick] ({#1-0.3},#2) to ({#1+0.3},#2);
    \draw[->] (#1+0.1,#2+0.25) to (#1+0.1,#2-0.25);
  \end{scope}
}
\newcommand{\orbDoubly}[2]
{
  \begin{scope}
    \draw[very thick] ({#1-0.3},#2) to ({#1+0.3},#2);
    \draw[->] (#1-0.1,#2-0.25) to (#1-0.1,#2+0.25);
    \draw[->] (#1+0.1,#2+0.25) to (#1+0.1,#2-0.25);
  \end{scope}
	}
\newcommand{\orbPartially}[2]
{
  \begin{scope}
    \draw[very thick] ({#1-0.3},#2) to ({#1+0.3},#2);
    \draw[->,dashed] (#1-0.1,#2-0.25) to (#1-0.1,#2+0.25);
    \draw[->,dashed] (#1+0.1,#2+0.25) to (#1+0.1,#2-0.25);
  \end{scope}
	}

%% file: orbs.tex

  \begin{tikzpicture}[scale=0.7]

  \tikzstyle{basis}=[draw, thick, rounded corners=0.1cm]
  \tikzstyle{CASSCF}=[basis, fill=blue!10];
  \tikzstyle{PFRS}=[basis, fill=blue!30];
  \tikzstyle{fix}=[basis, fill=orange!70];
  



 
  \draw[dashed,step=1.0cm] (6.75,3.5) -- (11.0,3.5);
  \draw[dashed,step=1.0cm] (6.75,5.7) -- (11.0,5.7);
  
  \draw[dashed,step=1.0cm] (6.75,0.5) -- (6.75,9.5);

  \node at (4,9.5) {\small \MO{} basis}; 
  \node at (8.5,9.5) {\small \NO{} basis}; 
  
 \node at (10,2.0) {\small ${\tilde i},{\tilde j}, ...$}; 
 \node at (10,4.5) {\small ${\tilde u},{\tilde v}, ...$}; 
 \node at (10,7.0) {\small ${\tilde a},{\tilde b}, ...$}; 
 \node [rotate=90] at (8.5,2) {Closed};
 \node [rotate=90] at (8.5,4.6) {Active};
 \node [rotate=90] at (8.5,7) {Virtual}; 



  \def\shift{3.0};  
  \node at (-1.0+\shift,8.7) {\small $|I\rangle$}; 
  \orbDoubly{-1.025+\shift}{1.0};
  \orbDoubly{-1.025+\shift}{1.7};
  \node at (-1.025+\shift,2.5) {$\vdots$};
  \orbDoubly{-1.025+\shift}{3.1};

  \orbDoubly{-1.025+\shift}{3.9};
  \orbEmpty{-1.025+\shift}{4.6};
  \orbEmpty{-1.025+\shift}{5.3};

  \orbEmpty{-1.025+\shift}{6.1};
  \orbEmpty{-1.025+\shift}{6.8};
  \node at (-1.025+\shift,7.5) {$\vdots$};
  \orbEmpty{-1.025+\shift}{8};

  \def\shift{5.0};  
  \node at (-1.0+\shift,8.7) {\small $|J\rangle$}; 
  \orbDoubly{-1.025+\shift}{1.0-0.1};
  \orbDoubly{-1.025+\shift}{1.7+0.1};
  \node at (-1.025+\shift,2.5) {$\vdots$};
  \orbDoubly{-1.025+\shift}{3.1-0.05};

  \orbEmpty{-1.025+\shift}{3.9+0.2};
  \orbDoubly{-1.025+\shift}{4.6-0.1};
  \orbEmpty{-1.025+\shift}{5.3+0.05};

  \orbEmpty{-1.025+\shift}{6.1-0.05};
  \orbEmpty{-1.025+\shift}{6.8-0.05};
  \node at (-1.025+\shift,7.6) {$\vdots$};
  \orbEmpty{-1.025+\shift}{8.1+0.05};

  \def\shift{7.0};
  \node at (-1.0+\shift,8.7) {\small $|K\rangle$}; 
  \orbDoubly{-1.025+\shift}{1.0+0.05};
  \orbDoubly{-1.025+\shift}{1.7-0.12};
  \node at (-1.025+\shift,2.4) {$\vdots$};
  \orbDoubly{-1.025+\shift}{3.1-0.2};

  \orbDown{-1.025+\shift}{3.9};
  \orbEmpty{-1.025+\shift}{4.6+0.05};
  \orbUp{-1.025+\shift}{5.3-0.25};

  \orbEmpty{-1.025+\shift}{6.1};
  \orbEmpty{-1.025+\shift}{6.8};
  \node at (-1.025+\shift,7.6) {$\vdots$};
  \orbEmpty{-1.025+\shift}{8.05};

  \def\shift{8.5};
  \orbDoubly{-1.025+\shift}{1.0};
  \orbDoubly{-1.025+\shift}{1.7};
  \node at (-1.025+\shift,2.5) {$\vdots$};
  \orbDoubly{-1.025+\shift}{3.1};

  \orbPartially{-1.025+\shift}{3.9};
  \orbPartially{-1.025+\shift}{4.6};
  \orbPartially{-1.025+\shift}{5.3};

  \orbEmpty{-1.025+\shift}{6.1};
  \orbEmpty{-1.025+\shift}{6.8};
  \node at (-1.025+\shift,7.6) {$\vdots$};
  \orbEmpty{-1.025+\shift}{8.2};

		\end{tikzpicture}